# How the electron-phonon coupling mechanism work in metal superconductor

Qiankai Yao[1,2]

[1]College of Science, Henan University of Technology, Zhengzhou450001, China
[2]School of physics and Engineering, Zhengzhou University, Zhengzhou450001, China

**Abstract** Superconductivity in some metals at low temperature is known to arise from an electron-phonon coupling mechanism. Such the mechanism enables an effective attraction to bind two mobile electrons together, and even form a kind of pairing system(called Cooper pair) to be physically responsible for superconductivity. But, is it possible by an analogy with the electrodynamics to describe the electron-phonon coupling as a resistivity-dependent attraction? Actually so, it will help us to explore a more operational quantum model for the formation of Cooper pair. In particularly, by the calculation of quantum state of Cooper pair, the explored model can provide a more explicit explanation for the fundamental properties of metal superconductor, and answer: 1) How the transition temperature of metal superconductor is determined? 2) Which metals can realize the superconducting transition at low temperature?

## 1. Introduction

In the BCS theory[1], superconductivity is attributed to a phonon-mediated attraction between mobile electrons near Fermi surface(called Fermi electrons). The attraction is sometimes referred to as a residual Coulomb interaction[2] that can glue Cooper pair together to cause superconductivity. Historically, the electron-phonon coupling concept was suggested by Fröhlich in 1950[3], and confirmed by the discovery of isotope effect[4]. The BCS theory led to a fundamental understanding of superconducting phenomena in momentum space, but it needs to give a more heuristic picture[5,6].

In this paper, we begin with a decision of Brown-like motion with a particle-field Lagrangian for mobile electrons in metal environment. This Lagrangian takes a mathematical form similar to the electromagnetic one, and the similarity allows us to adopt a unified framework to describe the electromagnetic and acoustic phenomena. By the definition, it can yield an effective attraction between Fermi electrons to be physically responsible for electronic pairing. The effective Hamiltonian of paired electrons and its Schrödinger solution are presented, which will directly derive the transition temperature of metal materials, including a physical criterion for their working as superconductors.

## 2. Acoustic interaction

**1) From damping motion to acoustic interaction** As well known that, the electric conduction of metal materials is achieved by means of the drift motion of large number mobile electrons. When conducting in metal, mobile electrons will suffer a damping collision with other objects, and thus cause a collision-dependent resistivity

$$\rho = \eta n_e e^2, \qquad \eta = \frac{m_e}{\tau} \qquad (1)$$

$n_e$ is the mobile electron density, $\eta$ the damping coefficient that is determined by the collision time $\tau$.

In metal environment, mobile electrons are usually modeled to be a kind of classical particles like gas molecules, each of which performs a Brown-like motion and satisfies the Langevin equation

$$m_e \frac{d\boldsymbol{v}}{dt} = -\eta \boldsymbol{v} - \nabla \kappa(\boldsymbol{r},t) \qquad (2)$$

Where, $\kappa$ denotes the random potential with an average equal to the thermal motion energy of mobile electrons(in natural units)

$$\bar{\kappa} = \frac{\pi^2 T^2}{4 E_F} \qquad (3)$$

$E_F$ is Fermi energy. Eq.(2) can be yielded by a particle-field Lagrangian

$$L(\boldsymbol{v},\boldsymbol{r},t) = \frac{1}{2} m_e v^2 - a \boldsymbol{A}_a(\boldsymbol{r},t) \cdot \boldsymbol{v} + \kappa(\boldsymbol{r},t) \qquad (4)$$

endowed with a differential relation

$$\frac{d\boldsymbol{A}_a}{dt} = \frac{\partial \boldsymbol{A}_a}{\partial t} + \boldsymbol{v} \cdot \nabla \boldsymbol{A}_a \qquad (5)$$

for the damping vector $\boldsymbol{A}_a = (\boldsymbol{v}t - \boldsymbol{r})\sqrt{4\pi\varepsilon_a}$, $a = \eta/\sqrt{4\pi\varepsilon_a}$. Although the presented Lagrangian only can lead to a Brown-like motion, rather than a bound state, it still gives us a firm base to develop the electronic pairing model.

From Lagrangian (4) we see that, the damping interaction is mathematically similar to the electromagnetic interaction. The similarity inspires us to treat $\boldsymbol{A}_a$ as an acoustic vector potential, and $a$ as an acoustic charge with mass density $\varepsilon_a$ playing the role of acoustic permittivity[7], analogous to the electric one $\varepsilon_e$. This means, a mobile electron in metal is actually acting as a dyon(quoted from Schwinger's invention[8]) of carrying dual charge $(-e, a)$, rather than a pure electrically charged particle. Importantly, the electromagnetic interaction essence of material elasticity and tension suggests, the acoustic equations should have the basic structure of electromagnetic equations. The suggestion allows us to formally write a joint electric-acoustic equation group[7]

$$\begin{cases} \nabla \cdot \mathbf{E} = \dfrac{\rho_e}{\varepsilon_e}, & \nabla \cdot \mathbf{B} = 0 \\ \nabla \times \mathbf{E} + \dfrac{\partial \mathbf{B}}{\partial t} = 0, & \nabla \times \mathbf{B} - \mu_e \varepsilon_e \dfrac{\partial \mathbf{E}}{\partial t} = \mu_e \boldsymbol{j}_e \\ \nabla \cdot \mathbf{b} = -\dfrac{\rho_a}{\varepsilon_a}, & \nabla \cdot \mathbf{e} = 0 \\ \nabla \times \mathbf{b} - \dfrac{\partial \mathbf{e}}{\partial t} = 0, & \nabla \times \mathbf{e} + \mu_a \varepsilon_a \dfrac{\partial \mathbf{b}}{\partial t} = \mu_a \boldsymbol{j}_a \end{cases} \quad (6)$$

$\rho_a$, $\boldsymbol{j}_a$ denote the acoustic charge and current densities, $\mu_a$ the reciprocal of elastic modulus(analogous to magnetic permeability $\mu_e$). Correspondingly, the related fields are defined by

$$\begin{cases} \mathbf{E} = -\nabla \varphi_e - \dfrac{\partial \boldsymbol{A}_e}{\partial t}, & \mathbf{B} = \nabla \times \boldsymbol{A}_e \\ \mathbf{b} = -\nabla \varphi_a - \dfrac{\partial \boldsymbol{A}_a}{\partial t}, & \mathbf{e} = -\nabla \times \boldsymbol{A}_a \end{cases} \quad (7)$$

The vibrating and shearing fields $\mathbf{b}$, $\mathbf{e}$ can be thought of as the quantum of acoustic wave in term of phonon, by which the acoustic interaction is mediated(analogous to the electro-magnetic interaction mediated by photon). When the medium condition of $1/\sqrt{\mu_a \varepsilon_a} \approx 1/\sqrt{\mu_e \varepsilon_e}$ is met, Eq.(6) can be amalgamated into a Heaviside-like form[9]

$$\begin{cases} \nabla \cdot \hat{\mathbf{E}} = \rho_e, & \nabla \times \hat{\mathbf{E}} + \sqrt{\mu_e \varepsilon_e} \dfrac{\partial \hat{\mathbf{B}}}{\partial t} = \boldsymbol{j}_a \\ \nabla \cdot \hat{\mathbf{B}} = -\rho_a, & \nabla \times \hat{\mathbf{B}} - \sqrt{\mu_e \varepsilon_e} \dfrac{\partial \hat{\mathbf{E}}}{\partial t} = \boldsymbol{j}_e \end{cases} \quad (8)$$

with

$$\begin{cases} \hat{\mathbf{E}} = \sqrt{\varepsilon_e}\mathbf{E} + \sqrt{1/\mu_a}\mathbf{e} \\ \hat{\mathbf{B}} = \sqrt{1/\mu_e}\mathbf{B} + \sqrt{\varepsilon_a}\mathbf{b} \end{cases}, \quad \begin{cases} \sqrt{\mu_e}\boldsymbol{j}_e \to \boldsymbol{j}_e, \ \sqrt{1/\varepsilon_e}\rho_e \to \rho_e \\ \sqrt{\mu_a}\boldsymbol{j}_a \to \boldsymbol{j}_a, \ \sqrt{1/\varepsilon_a}\rho_a \to \rho_a \end{cases} \quad (9)$$

In Eq.(8), the acoustic current $(\boldsymbol{j}_a, \rho_a)$ just occupies the position of the magnetic source. And hence, it is appropriate for us to phenomenologically understand the acoustic charge a as an acoustic "magnetic charge", and the vibrating(shearing) field $\mathbf{b}$ ($\mathbf{e}$) as an acoustic "magnetic (electric) field".

Not only that, with regard to Eq.(6), if introducing the following 2-component arrays

$$\begin{cases} (\bar{\boldsymbol{J}}, \bar{\rho}) = \left( \begin{bmatrix} \sqrt{\mu_e}\boldsymbol{J}_e \\ -\sqrt{\mu_a}\boldsymbol{J}_a \end{bmatrix}, \begin{bmatrix} \sqrt{1/\varepsilon_e}\rho_e \\ -\sqrt{1/\varepsilon_a}\rho_a \end{bmatrix} \right) \\ (\bar{\boldsymbol{A}}, \bar{\varphi}) = \left( \begin{bmatrix} \boldsymbol{A}_e/\sqrt{\mu_e} \\ \boldsymbol{A}_a/\sqrt{\mu_a} \end{bmatrix}, \begin{bmatrix} \sqrt{\varepsilon_e}\varphi_e \\ \sqrt{\varepsilon_a}\varphi_a \end{bmatrix} \right) \end{cases} \quad (10)$$

we also can equivalently express it in the Proca form[10]

$$\left( \nabla^2 - \dfrac{\partial^2}{\partial t^2} - \bar{\bar{\Lambda}} \right)(\bar{\boldsymbol{A}}, \bar{\varphi}) = -(\bar{\boldsymbol{J}}, \bar{\rho}), \quad \bar{\bar{\Lambda}} = \begin{bmatrix} \xi_e^{-2} & 0 \\ 0 & \xi_a^{-2} \end{bmatrix} \quad (11)$$

where, the acoustic interaction range(also called the coherent length) $\xi_a$ achieves its definition by following the electric screening length, namely

$$(\xi_e, \xi_a) = \left( \sqrt{\dfrac{\pi}{4m_e e^2 P_F}}, \sqrt{\dfrac{\pi}{4m_e a^2 P_F}} \right) \quad (12)$$
$$\sim (10^{-10}\,\mathrm{m}, 10^{-6}\,\mathrm{m})$$

$P_F$ is Fermi momentum. The two characteristic lengths respectively determine two modified frequent dispersions $\omega_{e,a}^2 = k_{e,a}^2 + 1/\xi_{e,a}^2$. It is tantamount to the photon and phonon obtaining their masses $1/\xi_{e,a}$. Eq.(11) provides us a physical framework to study the electron-phonon coupling. For example, by the equation we can write a composite Yukawa potential for a mobile electron with dual charge $(-e, a)$ at the zero point

$$\varphi_{(-e,a)}(r) = -\dfrac{e}{r}e^{-r/\xi_e} - \dfrac{a}{r}e^{-r/\xi_a} \quad (13)$$

It shows that, when the distance satisfies $\xi_e < r < \xi_a$, two acoustically charged electrons will produce a net Coulomb-like attractive force

$$F_{att}(r) \sim \dfrac{a^2}{r^2} = \dfrac{1}{4\pi\varepsilon_a n_e^2 e^4} \dfrac{\rho^2}{r^2} \quad (14)$$

The force is explicitly dependent on resistivity $\rho$ and mass density $\varepsilon_a$, that is, the bigger resistivity and smaller density, the stronger attraction.

Consequently, the electromagnetic analogy also allows us to express the acoustic force on an acoustically charged electron in a Maxwell-style form

$$\boldsymbol{F}_a = -\dfrac{a\partial \boldsymbol{A}_a}{\partial t} - a\nabla \varphi_a \quad (15)$$

Of which, the first term describes the damping effect, and the second the acoustic attraction. In the follows, we will present how the acoustic attraction can glue two Fermi electrons together to form a Cooper pair at low temperature.

**2) Cooper system** According to the BCS theory, two paired electrons must possess opposite momentum and spin[1]. However, in our model, opposite momentum implies opposite damping, and only the opposite damping can determine the zero resistance characteristic of Cooper system. So, in this sense, superconductivity doesn't exclude the existence of damping, but requires a zero damping effect on the whole.

In practice, once the attraction of acoustic charge is introduced, we can use Hamiltonian method to study electronic pairing. To this end, let us consider two acoustically charged Fermi electrons with opposite momentum and spin, which separated by a distance $2r(<\xi_a)$, only involve radial motion relative to their mass centre. Then, by canonical momentum

$$\boldsymbol{P} = \dfrac{\partial L}{\partial \dot{\boldsymbol{v}}} = \boldsymbol{p} - a_F \boldsymbol{A}_a \quad (16)$$

we have the following paired Hamiltonian

$$H(P_j, r_j) = \sum_{j=1}^{2} \left( \dfrac{P_j^2}{2m_e} - \dfrac{a_F^2}{4r_j} + \bar{\kappa}_{Fj} \right) \quad (17)$$

$a_F$ is the effective acoustic charge carried by Fermi electrons (shorter collision time determines $a_F \gg a$), $\bar{\kappa}_F$ the average thermal motion energy of Fermi electrons. In operator form, the Hamiltonian yields a s-wave symmetrical Schrödinger solution

$$\Psi(r_1, r_2) \sim e^{-\frac{r_1+r_2}{a_0}}, \qquad a_0 = \frac{4}{m_e a_F^2} \tag{18}$$

followed by a single-particle bound energy, i.e. disassembly energy

$$E(T) = E_0 - \bar{\kappa}_F(T), \qquad E_0 = \frac{m_e a_F^4}{32} \tag{19}$$

$E_0$ is the single-particle paired energy. Then, by identifying $\bar{\kappa}_F$ as

$$\bar{\kappa}_F = \frac{R_1^2 \bar{\kappa}}{\pi^2} \tag{20}$$

we can determine a critical temperature at which the bound energy is zero

$$T_c = \frac{2\Delta_0}{R_1}, \qquad \Delta_0 = \sqrt{E_F E_0} \sim 10^4 E_0 \tag{21}$$

$R_1$ is a scale parameter. The result shows, two Fermi electrons can be formed into a Cooper system only $T < T_c$, i.e. $\bar{\kappa}_F(T) < E_0$. **Fig.**1 illustrates the internal coupling pattern of Cooper system in metal, which has a characteristic radius of about $a_0 \sim 10^{-7}$ m, just like a "super parahelium atom".

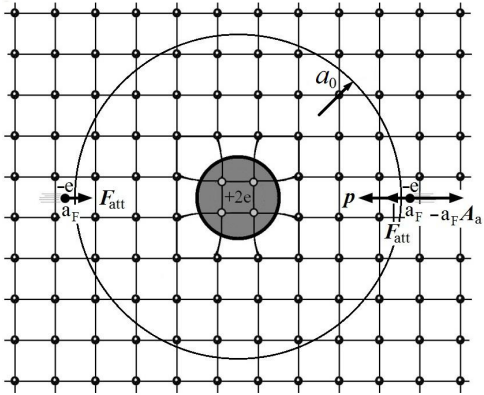

**Fig.**1 In the acoustic field of being characterized by lattice distortion, the Coulomb expression of acoustic potential $-2e^2/\varepsilon_{er} r$ can make a Cooper system look like a super parahelium atom of central charge $+2e$ with a relative permittivity $\varepsilon_{er} = 8e^2/a_F^2$.

In electromagnetic medium, there is usually a no-dissipation current to be produced through polarization (magnetization) process. To use this no-dissipation current to model superconductivity, we treat metal superconductor as a special medium of containing a large number of "super-atoms"(with a density $n_s$). As the medium polarized (magnetized), it will produce a polarized(magnetized) super-current $(j_s, \rho_s)$, namely

$$\begin{cases} j_s = \frac{\partial P_s}{\partial t} = \nabla \times M_s \\ \rho_s = \nabla \cdot P_s \end{cases}, \qquad \begin{cases} \nabla \cdot j_s = 0 \\ \frac{\partial \rho_s}{\partial t} = 0 \end{cases} \tag{22}$$

with the effective polarization and magnetization defined by

$$P_s = n_s p_s, \qquad M_s = n_s m_s \tag{23}$$

$p_s$, $m_s$ denote the induced dipole and magnetic moments of a super-atom. According to Cooper's suggestion[11], the induced supercurrent can be used to replace the electromagnetic potential to describe the electromagnetic properties of superconductor. For this purpose, we now rewrite the usual London equations

$$\begin{cases} \mathbf{E} = -\lambdabar^2 \frac{\partial j_s}{\partial t} \\ \mathbf{B} = -\lambdabar^2 \nabla \times j_s \end{cases}, \qquad \lambdabar = \sqrt{\frac{m_e}{2n_s e^2}} \tag{24}$$

in a generalized form

$$\begin{cases} \mathbf{E} = -\lambdabar^2 \left(\nabla \rho_s + \frac{\partial j_s}{\partial t}\right) \\ \mathbf{B} = -\lambdabar^2 \nabla \times j_s \end{cases} \tag{25}$$

Then, by the generalized we can deduce a modified Proca equation with a normal current $(j_n, \rho_n)$ as its source, that is

$$\left(\nabla^2 - \frac{\partial^2}{\partial t^2} - \frac{1}{\lambdabar^2}\right)(j_s, \rho_s) = \frac{1}{\lambdabar^2}(j_n, \rho_n) \tag{26}$$

Since $\partial \rho_s / \partial t = 0$ and $\rho_s + \rho_n = 0$ (determined by the electric neutrality of superconductor), the density component of Eq.(26) yields $\nabla \rho_s = 0$. This is very the condition that the usual London equations need. In the stationary case($\partial/\partial t \to 0$ and $j_n = 0$), the other transforms into the Poisson's expression

$$\nabla^2 j_s = \frac{1}{\lambdabar^2} j_s \tag{27}$$

The equation has a decreasing solution $j_s = j_{s0} \exp(-x/\lambdabar)$, whose falling scale $\lambdabar$ represents a measure of the depth to which the induced supercurrent can penetrate the super-conducting body. Like the superconducting coherent length $\xi_a$, the penetration depth $\lambdabar$ is also a characteristic parameter of superconductor.

## 3. Explanation for superconducting properties

**1) Heat capacity** By substituting kinetic momentum $p$ for $P$ in Hamiltonian (17), we can calculate its single-particle state average

$$\langle H(p,r) \rangle = (E_F - E) - i\Delta, \qquad \Delta = \frac{a_F}{2m_e} \langle \partial_r A_a \rangle \tag{28}$$

with an imaginary part $\Delta$ serving as the energy gap. Here, the average of damping momentum is identified as $a_F \langle A_a \rangle = P_F$, only so, it can be compatible with the bound state of paired electrons. Accordingly, the condensed energy is found by

$$E_c = |E_F - \langle H(p,r) \rangle| = \sqrt{E^2 + \Delta^2} \tag{29}$$

When temperature $T < T_c$, there would be a considerable amount of paired electrons entering a collective state, and immediately the metal body itself undergoes a superconducting transition. As shown in **Fig**.2, the collective state is separated by the condensed energy from the higher states, and so have the excitation spectrum

$$E_{\text{exc}} = |E_p - \langle H(\bm{p},\bm{r})\rangle| = \sqrt{(E_p - E_F + E)^2 + \Delta^2} \quad (30)$$

$E_p$ is the kinetic energy of mobile electron. In the case of neglecting the bound energy $E(\ll \Delta \ll E_F)$, Eq.(30) reduces to the usual form. The result means, although entering a condensed state, the paired electrons are still involved in Fermi statistics, but leave a vacant band—energy gap in momentum space(see **Fig**.2).

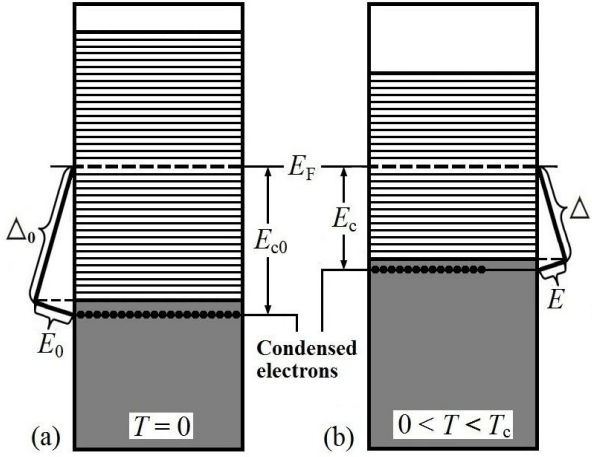

**Fig**.2 The level arrangements of metal superconductor at absolute zero and finite temperatures are plotted by (a), (b), including an energy gap $\Delta$ left by paired electrons as they all collect into a level of $E_c(\approx \Delta)$ below Fermi surface.

Notice that, the density $n_s$ varies between $n_0$ and $0$, it is the BCS microscopic justification for the two-fluid model of superconductivity: the normal fluid consists of the thermally excited quasi-particle gas, the superfluid is the condensate of Cooper pairs. Crucially, as a vacant band left by paired electrons, the gap width undoubtedly has the meaning of energy falling, and thus statistically related to the normal fluid density $n_n(T)$ by Boltzmann distribution. This decision can help us to define a dimensionless function

$$Y(T) = \frac{n_n}{n_0} = \Theta e^{-\Delta_0/T} \quad (31)$$

with a coefficient $\Theta$ fixed by the self-consistent condition: $Y(T \to T_c) \to 1$. $Y(T)$ is called Yoshida function[12], whose variation is reflected in thermodynamic quantities as an exponentially activated behavior at low temperature, due to the small number of quasi-particles with thermal energy sufficient to be excited over it.

The electronic specific heat is easy to calculate, since the entropy of superconductor is once again the entropy of free gas of electronic excitons. Being statistical, we can deduce the heat absorbed by a superconducting system in certain process should be proportional to Yoshida function and temperature interval

$$dQ = CY(T)dT \quad (32)$$

$C$ is a scale coefficient. Correspondingly, the total heat released by the system in the process of temperature $T_c \to 0$, should be the sum of the total thermal and condensed energies

$$\int_0^{T_c} dQ(T) \approx C\left(0.1T + \frac{0.2T^2}{T_c}\right)Y(T)\bigg|_0^{T_c} \quad (33)$$
$$= n_e\bar{\kappa}(T_c) + n_0\Delta_0$$

If noting a well-known ratio $n_e\bar{\kappa}(T_c) : n_0\Delta_0 = 2:1$ given by the two-fluid model[13], we clarify

$$R_1 = \frac{2\pi}{\sqrt{3}}, \qquad C = 2.50 C_n(T_c) \quad (34)$$

$C_n$ is the normal specific heat. And so, the superconducting specific heat reads

$$C_s(T) = \frac{dQ(T)}{dT} = 2.50 C_n(T_c)Y(T) \quad (35)$$

Following it is a normalized discontinuity at the transition point

$$R_2 = \frac{C_s - C_n}{C_n}\bigg|_{T=T_c} = 1.50 \quad (36)$$

The variation tendency is identical to the measurements[14].

Moreover, the presented energy level also tells us, the differential heat of a superconductor should be proportional to the variation of normal fluid, namely

$$dQ \propto \Delta dn_n \propto -\Delta d\Delta \quad (37)$$

Then, by the differential relation and Eq.(33), we get a temperature-dependent gap approximately

$$\Delta(T) \approx \Delta_0\sqrt{1 - \frac{1}{3}\left(\frac{T}{T_c} + \frac{2T^2}{T_c^2}\right)Y(T)}$$
$$\approx \begin{cases} \Delta_0\left(1 - \frac{1.02T}{T_c}e^{-\Delta_0/T}\right), & T \ll T_c \\ 1.35\Delta_0\sqrt{1 - \frac{T}{T_c}}, & T \sim T_c \end{cases} \quad (38)$$

The way in which the reduced gap varies with temperature is plotted in **Fig**.3.

**2) Transition temperature** Now, it can be asserted that, the paired energy $E_0$ is eventually linked with the metallic resistivity(through damping coefficient $\eta$), and the related resistivity should refer to the effective residual component $\rho_r$, rather than the usual one. So that, using the expansion with $T^2$, we have the following Matthiessen form

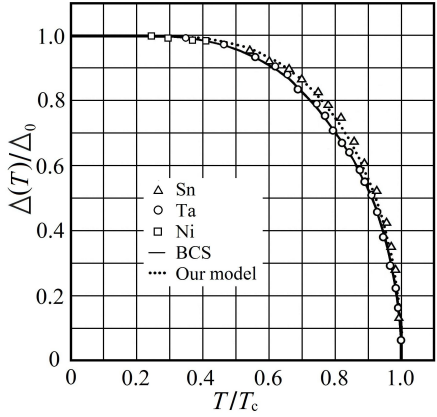

**Fig.**3 The variation of reduced gap with temperature. The full curve shows the BCS prediction, and the dotted is given by Eq.(38). The experimental points are those found for metals Sn, Ta and Ni[15].

$$\rho(T^2) \approx \rho_r + \rho_T T^2 \quad (39)$$

corresponding to a linear expression near a designated temperature $T_0$, namely

$$\rho = \rho_0[1 + \alpha(T - T_0)] \quad (40)$$

$\alpha$ represents temperature coefficient. The consistency of two expressions at $T_0$ requires

$$\rho_r + \rho_T T_0^2 = \rho_0, \quad 2\rho_T T_0 = \alpha\rho_0 \quad (41)$$

giving the residual resistivity

$$\rho_r = \rho_0\left(1 - \frac{\alpha T_0}{2}\right) \quad (42)$$

Substituting it into the expression of charge $a(\eta(\rho_r))$, we have

$$a = \frac{\rho_0(2 - \alpha T_0)}{4n_e e^2 \sqrt{\pi \varepsilon_a}} \quad (43)$$

To deduce the superconducting transition temperature, we now investigate the acoustic interaction from an average point of view, that is, the general electrons of carrying charge $a$ also have a pairing propensity. This consideration allows us to follow the gap $\Delta_0$ to write an average energy falling

$$\overline{\Delta}_0 = \sqrt{\frac{m_e a^4 E_F}{32}} \quad (44)$$

which is identical to the Cooper condensation only

$$\overline{\Delta}_0 = \frac{n_0}{n_e}\Delta_0 = \frac{3\Delta_0^2}{8E_F} \quad (45)$$

Then, by the last three relations and Eq.(21) we get

$$T_c = \frac{ne^2 \rho_0(2 - \alpha T_0)\sqrt{3\delta^3}}{4\sqrt{2\pi m_e M}} \quad (46)$$

here, $\varepsilon_a = nM$ and $n_e = \delta n$ are used, $\delta$, $n$ denote the valence and the atom density, M the atomic weight. Eq.(46) directly leads to isotope effect $T_c \propto M^{-1/2}$, whose resistivity-dependence could explain a paradox that those elements that are the best conductors of electricity(Cu, Ag and Au) do not become superconductors even at very low temperature, while poorer conductors(like Tc, La and Pb) have a higher transition temperature[16]. **Table**1 gives the transition temperatures of some metal superconductors, except a few of materials(such as Ti, Zr, Nb, Hf and W), the theoretical calculations($T_0 = 293$K ) are roughly in agreement with the experimental results(the status can be improved, if selecting more appropriate reference temperature).

**Table**1 The comparison between theoretical calculations(t) and experimental results(e) for transition temperature of some metal superconductors( $\rho_0(10^{-8}\Omega\cdot m)$, $\alpha(10^{-3}K^{-1})$, $T(K)$ )(* $T_0 = 161$K for Hg).

| Element($\delta$) | $\rho_0$ | $\alpha$ | $\sigma$ | $T_c(t)$ | $T_c(e)$ | Element($\delta$) | $\rho_0$ | $\alpha$ | $\sigma$ | $T_c(t)$ | $T_c(e)$ |
|---|---|---|---|---|---|---|---|---|---|---|---|
| Al(3) | 2.8 | 4.4 | 0.93 | 1.01 | 1.14 | Sn(4) | 12.6 | 4.5 | 1.43 | 1.55 | 3.72 |
| Ti(2) | 42.0 | 4.1 | 2.99 | 6.56 | 0.39 | La(2) | 57.0 | 2.6 | 3.33 | 3.87 | 6.00 |
| V(2) | 19.6 | 3.9 | 2.09 | 4.06 | 5.38 | Hf(2) | 32.2 | 4.4 | 1.78 | 1.86 | 0.12 |
| Zn(2) | 5.2 | 4.2 | 0.96 | 0.78 | 0.88 | Ta(2) | 19.0 | 3.5 | 1.59 | 1.83 | 4.48 |
| Ga(3) | 13.6 | 5.0 | 1.40 | 1.95 | 1.09 | W(2) | 5.5 | 4.5 | 0.71 | 0.42 | 0.01 |
| Zr(2) | 42.0 | 4.4 | 2.40 | 3.21 | 0.55 | Re(2) | 18.5 | 4.5 | 1.30 | 1.50 | 1.40 |
| Nb(1) | 16.0 | 2.6 | 1.64 | 0.97 | 9.50 | Os(2) | 8.8 | 4.2 | 0.95 | 0.84 | 0.66 |
| Mo(1) | 5.7 | 3.3 | 0.89 | 0.33 | 0.92 | Ir(2) | 6.5 | 3.9 | 0.86 | 0.68 | 0.14 |
| Tc(2) | 20.0 | 3.7 | 1.85 | 3.12 | 7.77 | Hg(2)* | 49.0 | 6.1 | 2.56 | 3.62 | 4.15 |
| Ru(1) | 7.3 | 4.1 | 0.87 | 0.36 | 0.51 | Tl(3) | 19.0 | 5.0 | 1.27 | 1.09 | 2.39 |
| Cd(2) | 7.3 | 4.3 | 0.97 | 0.56 | 0.56 | Pb(4) | 21.0 | 3.7 | 1.87 | 2.99 | 7.19 |
| In(3) | 8.8 | 5.1 | 0.97 | 0.70 | 3.40 | Th(2) | 16.0 | 2.4 | 1.59 | 0.99 | 1.37 |

**3) Superconducting index** Regarding the condition of superconducting transition, we from the temperature-dependence of bound energy find, no matter how weak the acoustic interaction is, as long as the temperature low enough, there will eventually be Cooper pairs being formed. This seems to imply, at sufficiently low temperature, any metal could become a superconductor, but that is not the case. The reason is that, in order to prevent Cooper body from being torn by shearing motion of supercurrent, the penetration depth $\lambda$ cannot be too short. On the other hand, in order to repel magnetic field as far as possible, $\lambda$ should also be short enough(see **Fig**.4). Such opinion can help us to use the zero-temperature penetration depth $\lambda_0$ to define a superconducting index for metal materials, that is

$$\sigma = \frac{\lambda_0}{a_0} = \left(\frac{2\pi\delta m_e^3}{M}\right)^{1/4} \sqrt{\frac{\rho_0(2-\alpha T_0)}{3}} \qquad (47)$$

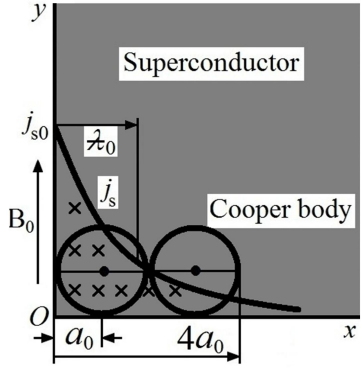

**Fig**.4 The penetration depth of supercurrent $j_s$ (along × direction) is always restricted by Cooper body.

As shown in **Table**1, a metal(excluding W) can become a superconductor at low temperature only its superconducting index $\sigma \in [0.86, 3.33]$, or, it will be a non-superconductor(see **Table**2).

**Table**2 Superconducting indexes of some alkali, ferromagnetic, precious and lanthanide metals($^\# T_0 = 244\,\text{K}$ for Pt).

| Element($\delta$) | $\rho_0$ | $\alpha$ | $\sigma$ | Element($\delta$) | $\rho_0$ | $\alpha$ | $\sigma$ |
|---|---|---|---|---|---|---|---|
| Na(1) | 4.7 | 5.5 | 0.71 | Pt(1)$^\#$ | 8.6 | 4.5 | 0.85 |
| Mg(2) | 4.2 | >6.8 | 0.00 | Au(1) | 1.2 | 4.0 | 0.41 |
| K(1)  | 6.8 | 5.7 | 0.69 | Pr(2) | 68.0 | 1.7 | 3.99 |
| Ca(2) | 3.7 | 4.6 | 0.84 | Nd(2) | 64.0 | 1.6 | 3.88 |
| Fe(2) | 10.1 | 6.5 | 0.49 | Sm(2) | 92.0 | 1.8 | 4.52 |
| Co(2) | 6.3 | 6.6 | 0.32 | Gd(2) | 134.0 | 0.9 | 5.86 |
| Ni(1) | 6.9 | 6.8 | 0.11 | Tb(2) | 116.0 | 1.5 | 5.15 |
| Cu(1) | 1.7 | 4.3 | 0.45 | Dy(2) | 91.0 | 1.2 | 4.67 |
| Rb(1) | 12.1 | 5.5 | 0.82 | Ho(2) | 94.0 | 1.7 | 4.51 |
| Pd(0) | 10.8 | 4.2 | 0.00 | Er(2) | 86.0 | 2.0 | 4.17 |
| Ag(1) | 1.6 | 4.1 | 0.40 | Tm(2) | 90.0 | 2.0 | 4.25 |

**4) Magnetism** An important nature of superconductor is that superconductivity can be quenched by magnetic field. To illustrate this, let us imagine there is a magnetized super-atom interacting with field **B** through its induced moment $m_s$, whose interaction energy—magnetic energy can be decomposed along two components of condensed energy, that is

$$m_s \text{B} \to \frac{m_s \text{B}}{E_{c0}}(E_0 + i\Delta_0) \qquad (48)$$

So that, once the magnetic energy exceeds the bound energy of super-atom $2E_c$, it will first use its real part to disassemble the super-atom, and then excite two disassembled electrons into the normal states. Physically, such a decision can provide an inequality $m_s \text{B}|^{\text{Re}} \geq 2E$, by which we deduce the critical field

$$\text{B}_c(T) = \text{B}_0\left(1 - \frac{T^2}{T_c^2}\right), \qquad \text{B}_0 = \frac{2E_{c0}}{m_{s0}} \approx \frac{2\Delta_0}{m_{s0}} \qquad (49)$$

The experimental results show this is indeed so[17].

Therefore, if $F_n$ is the free energy of a metal in normal state, $F_s$ the free energy in superconducting state in zero field, the difference between them should be equal to the critical magnetic work

$$F_n - F_s = \int_0^{B_c} M_s \text{dB} = n_0 \Delta_0 \left(1 - \frac{T^2}{T_c^2}\right)^2 \qquad (50)$$

This is the condensed energy of superconducting body. Different form the procedure of Ginzburg-Landau theory[18], the result is obtained by examining the interaction between the applied field **B** and the induced moment $m_s$, which gives a ratio

$$R_3 = \frac{\gamma T_c^2}{4\pi \text{B}_0^2} = \frac{1}{2\pi} \qquad (51)$$

$\gamma$ is Sommerfeld constant. By the comparison of theoretical and experimental results[19], three ratios $R_1$, $R_2$ and $R_3$ for some superconductors, are shown in **Table**3.

**Table**3. Comparison between theoretical calculations(t) and experimental measurements(e) for several indicators of some superconductors.

| Element | $2\Delta_0$(t) | $2\Delta_0$(e) | $\text{B}_0$(t) | $\text{B}_0$(e) | $R_1$ | $R_2$ | $R_3$ |
|---|---|---|---|---|---|---|---|
| Al | 3.2 | 3.4 | 7.7 | 10.5 | 3.3 | 1.45 | 0.171 |
| V  | 12.7 | 16.0 | 29.6 | 142.0 | 3.4 | 1.57 | 0.170 |
| Zn | 2.4 | 2.4 | 5.6 | 5.3 | 3.2 | 1.30 | 0.177 |
| Ga | 6.1 | 3.3 | 14.3 | 5.1 | 3.5 | 1.44 | 0.169 |
| Nb | 3.0 | 30.5 | 6.0 | 198.0 | 3.80 | 1.87 | 0.157 |
| Mo | 1.0 | 2.7 | 2.1 | 9.5 | 3.4 | 1.28 | 0.182 |
| Cd | 1.8 | 1.5 | 3.8 | 3.0 | 3.2 | 1.32 | 0.177 |
| In | 2.2 | 10.5 | 4.9 | 29.3 | 3.6 | 1.73 | 0.157 |
| Sn | 4.9 | 11.5 | 10.9 | 30.9 | 3.5 | 1.60 | 0.161 |
| La | 12.1 | 19.0 | 23.9 | 110.0 | 3.7 | 1.50 | — |
| Ta | 5.7 | 14.0 | 12.7 | 83.0 | 3.60 | 1.59 | 0.161 |
| Hg | 11.3 | 16.5 | 24.1 | 41.2 | 4.6 | 2.18 | 0.134 |
| Tl | 3.4 | 7.4 | 7.6 | 17.1 | 3.57 | 1.50 | 0.161 |
| Pb | 9.3 | 27.3 | 21.4 | 80.3 | 4.38 | 2.65 | 0.134 |
| BCS theory | | | | | 3.53 | 1.43 | 0.168 |
| Acoustic charge model | | | | | 3.62 | 1.50 | 0.159 |

Finally, the experiment also shows, whether a superconductor is put into rotation or a rotating normal metal is cooled into the superconducting state, it will develop a uniform parallel and rotation-determined magnetic field

$$\mathbf{B}_{\text{rot}} = \frac{2m_e \hat{\omega}}{e} \qquad (52)$$

throughout its interior[20]. The induction of magnetic field implies, the paired electrons must speed up when a superconductor starts to rotate. However, why? According to our model, a rotating superconductor can cause a normal current

$$j_n = 2n_s e r\widehat{\omega} n_\tau \quad (53)$$

$n_\tau$ is the unit tangent vector. The normal current by Eq.(26) determines a supercurrent $j_s = -j_n$. Serving as an electro-magnetic potential, the supercurrent $j_s$ by Eq.(24) naturally induces the rotation-dependent field $B_{rot}$. That is to say, the defining property of superconductors to expel any magnetic field from its bulk reverses when it is rotated.

### 4. Summary

In comparison with the BCS theory, the developed model has a conceptual difference in addition to those already described. The BCS theory emphasizes, the paired electrons are bound by mutual exchange of virtual phonons. Our model describes this exchange mechanism as a resistivity-dependent attraction that acts as a "glue" to bind Cooper pairs together.

In BCS picture, the condensed energy depends strongly on both the coupling strength and the cutoff frequency, but being a non-analytic function of involved parameters. Therefore, it is difficult to arrive at the result at any order in perturbation theory. In contrast, due to removing the complexity of BCS theory, the mentioned essential singularity does not appear in our model.

The BCS theory cannot offer more detailed information about Cooper system. The developed not only can calculate its quantum state and level structure, but also lead to the transition temperature of metal superconductors. Specifically, the introduction of superconducting index brings us a criterion to decide which metals can enter superconducting state.

Reviewing the overall scenario and its implications, the most remarkable point is that, our model can naturally lead to the fundamental behaviors of metal superconductors, such as the condensed energy, Meissner effect, isotope effect, excitation spectrum, specific heat and magnetism. Yet, it requires no more assumptions to match the experimental results. Although being specially designed for metal superconductors, the electron-phonon coupling mechanism still tells us a basic physics, that is, in order to transport charge without hindrance, it is necessary for superelectrons to enter a special(mobile and bound) quantum state, in which they would be endowed with an ability of avoiding dissipation in migrating process. This ability implies, in all the superconducting materials(both conventional and non-conventional), there must exist an interaction that can either tie mobile electrons together or give the original bound electrons a conductivity. The situation determines that the production of superelectrons is nothing more than two disparate possible ways: one is to make mobile electrons into bound quantum state(it is very the working principle of conventional superconductors), the other to allow the original bound electrons move freely(it is an alternative to non-conventional superconductors). Maybe the opinion provides a guideline in the search for high temperature superconductors.

Email: yaoqk@zzu.edu.cn